\documentclass[aps,pre,floats,twocolumn,showpacs,superscriptaddress]{revtex4}

\usepackage{epsfig}
\usepackage{latexsym,amsmath}

\begin{document}
\title{Clustering in complex networks. I. General formalism}

\author{M. \'Angeles Serrano}

\affiliation{School of Informatics, Indiana University,\\ Eigenmann
Hall, 1900 East Tenth Street, Bloomington, IN 47406, USA}

\author{Mari{\'a}n Bogu{\~n}{\'a}}

\affiliation{Departament de F{\'\i}sica Fonamental, Universitat de
  Barcelona,\\ Mart\'{\i} i Franqu\`es 1, 08028 Barcelona, Spain}

\date{\today}

\begin{abstract}
We develop a full theoretical approach to clustering in complex
networks. A key concept is introduced, the edge multiplicity, that
measures the number of triangles passing through an edge. This
quantity extends the clustering coefficient in that it involves the
properties of two --and not just one-- vertices. The formalism is
completed with the definition of a three-vertex correlation
function, which is the fundamental quantity describing the
properties of clustered networks. The formalism suggests new metrics
that are able to thoroughly characterize transitive relations. A
rigorous analysis of several real networks, which makes use of the
new formalism and the new metrics, is also provided. It is also
found that clustered networks can be classified into two main
groups: the {\it weak} and the {\it strong transitivity} classes. In
the first class, edge multiplicity is small, with triangles being
disjoint. In the second class, edge multiplicity is high and so
triangles share many edges. As we shall see in the following paper,
the class a network belongs to has strong implications in its
percolation properties.

\end{abstract}


\maketitle

\section{Introduction}
\label{Sec:Intro}
The important role of transitive relations in complex interaction
systems has been exposed since the work of Georg Simmel, a popular
19th century German sociologist who pointed out the interest in
triads in a pioneering work on the concept of social
structure~\cite{Simmel:1908}. Simmel understood society as a web of
patterned interactions  and focused on the study of the forms of
these interactions as they occur and reoccur in diverse historical
periods and cultural settings. His emphasis on quantitative aspects
lead him to analyze, in particular, dyadic versus triadic
relationships, to find that when a dyad is formed into a triad, the
apparently insignificant fact that one member has been added
actually brings about a major qualitative change, various actions
and processes becoming possible where previously they could not take
place. The triad is then seen as the simplest structure in which the
group as a whole can achieve domination over its component members,
and so becomes the scenario exhibiting the simplest expression of the
sociological drama.

In the study of complex networks, where large systems of
interactions are mapped into comprehensible
graphs~\cite{Albert:2002,Dorogovtsev:2003,Newman:2003}, just
vertices and edges are nevertheless usually recognized as the
primary building blocks. Vertices represent the elementary units
under mutual influence, and the interactions are modeled by edges
linking them. Transitive relations, represented in this scheme by
triangles, arise then as a secondary form of basic organization,
made up of vertices connected by edges. However, the empirical
evidence of a big number of triangles well above random expectations
in the vast majority of real networks has brought this figure into
attention, with a first reference to transitivity appearing in the
literature of complex networks in the form of the clustering
coefficient~\cite{Watts:1998}, a scalar measure quantifying the
total number of triangles in a network through the average likelihood
that two neighbors of a vertex are neighbors themselves. Triangles
in complex networks are indissolubly tied to the analysis of degree
correlations and they have been recognized as a fundamental element
in the composition of recurring subgraphs, the so-called
motifs~\cite{Milo:2002}, closely related to the large-scale
organization of complex networks~\cite{Vazquez:2004}, their
functionality or community
structure~\cite{Radicchi:2004,Palla:2005}. So, in the framework of
complex networks science, they have to be taken into account as a
basic unbridged object, whose presence and self-organization can
drastically impact network structure and properties.

In this paper and the following one, we develop a full theoretical
approach to clustering in complex networks on the basis of former
work~\cite{Serrano:2005b,Serrano:2006}. It is extended and completed
with novel results previously unreported which lead to a
substantially improved understanding of how clustering can be
measured and which is the reach of its effects. In this paper, we
begin by exposing in the next three sections the ways of measuring
clustering at different depth levels. In section~\ref{Sec:ck}, we
review, as a technical introduction and for completeness, the
standard local and global measures related to one-vertex clustering.
In section~\ref{Sec:mkk}, we ask for the properties of not just one
but two of the vertices involved in the triangles and to this end we
review the concept of dyadic clustering from the definitions of edge
multiplicity and edge clustering. Section~\ref{Sec:Q} treats the
case of triadic clustering. In particular, we propose a new measure,
the average nearest neighbors multiplicity $\bar{m}_{nn}(k)$, to
compute triadic clustering in a practical way. In
Section~\ref{Sec:corr}, we explore the effects of degree-degree
correlations on clustering. We find analytically that degree-degree
correlations constrain the functional form of clustering and its
maximum level. We also examine some empirical networks, finding a
good agreement with our predictions. Section~\ref{Sec:uncorr}
explores the condition for the simultaneous absence of degree
correlation at the level of triangles and edges, which makes
necessary the discrimination between weak and strong clustering.
Finally, conclusions are drawn in Section~\ref{Sec:conclusions}. In this way,
this first paper lays the general formalism. The following one will
focus on percolation properties.

\section{One-vertex clustering}
\label{Sec:ck}
In the context of complex networks, the concept of clustering was
introduced as a way to quantify the transitivity of the connections.
Several alternative definitions have been proposed, from global
scalar quantities associated to the whole
network~\cite{Barrat:2000,Newman:2001b} to local measures describing
the properties of single nodes. This is the case of the clustering
coefficient first introduced by Watts and
Strogatz~\cite{Watts:1998},
\begin{equation}
c_i=\frac{2 T_i}{k_i (k_{i} -1)},
\end{equation}
where $T_i$ is the number of triangles passing through vertex $i$
and $k_{i}$ is its degree. They also pointed out that real networks
display a level of clustering --measured as the average of $c_i$
over the set of vertices in the network, the so-called clustering
coefficient $C$ varying in the interval $[0,1]$-- typically much
larger than that produced by random effects.

The local clustering $c_{i}$ gives highly detailed information from
a purely local perspective. One can adopt a compromise between the
global property defined by $C$ and the full local information given
by $c_{i}$ by defining an average of $c_{i}$ over the set of
vertices of a given degree class~\cite{Vazquez:2002}, that is,
\begin{equation}
\bar{c}(k)=\frac{1}{N_k} \sum_{i \in \Upsilon(k)} c_{i}=\frac{1}{k
(k-1)N_k} \sum_{i \in \Upsilon(k)} 2T_{i} ,
\label{c(k)}
\end{equation}
where $N_{k}$ is the number of vertices of degree $k$ and
$\Upsilon(k)$ is the set of such vertices. The corresponding scalar
measure is called the mean clustering coefficient and can be computed on the
basis of the degree distribution $P(k)$ as
\begin{equation}
  \bar{c} = \sum_{k} P(k) \bar{c}(k),
  \label{eq:30}
\end{equation}
which is related to the clustering coefficient as
$C=\bar{c}/(1-P(0)-P(1))$. In fact, we have implicitly assumed that
$\bar{c}(k=0)=\bar{c}(k=1)=0$, whereas in the definition of $C$ we
only consider an average over the set of vertices with degree $k >
1$. This fact explains the difference between both measures.

In the case of uncorrelated networks, $\bar{c}(k)$ is independent of
$k$. Furthermore, all the measures collapse and reduce to
$C$~\cite{Newman:2003,Boguna:2003b,Burda:2004}.
\begin{equation}
  \bar{c}(k) = C = \frac{1}{N}\frac{(\langle k^2 \rangle-\langle k \rangle)^2}{\langle k \rangle^3} \mbox{  ,  } k > 1.
  \label{eq:32}
\end{equation}
Therefore, a functional dependence of $\bar{c}(k)$ on the degree can
be attributed to the presence of correlations. Indeed, it has been
observed that many real networks exhibit a power-law behavior
$\bar{c}(k)\sim k^{-\alpha}$, with typically $0\le \alpha \le 1$. Hence, the degree dependent clustering coefficient has been proposed as a measure of
hierarchical organization and modularity in complex
networks~\cite{Ravasz:2003}.

\section{Dyadic clustering}
\label{Sec:mkk}
The degree dependent clustering coefficient described in the
previous section measures the transitivity of a vertex that
participates in a triangle and, in this sense, it is a projection
over one vertex of a structure that involves three vertices. Then,
it is natural to ask for the properties of not just one but two of
the vertices involved in the triangle or, equivalently, to ask for the properties of edges involved in triangles.

To do so, let us define the multiplicity of an edge, $m_{ij}$, as
the number of triangles in which the edge connecting vertices $i$
and $j$ participates. This quantity is the analog for edges to the
number of triangles attached to a vertex, $T_{i}$. The two
quantities are related through the trivial identity
\begin{equation}
\sum_{j} m_{ij}a_{ij} = 2 T_i,
\label{trivialident}
\end{equation}
which is valid for any network configuration. The matrix $a_{ij}$ is
the adjacency matrix, giving the value $1$ if there is an edge
between vertices $i$ and $j$ and $0$ otherwise.

Again, $m_{ij}$ is a local measure defined for every edge. We can coarse-grain and define the average multiplicity of the edges connecting the
degree classes $k$ and $k'$, $m_{kk'}$, as
\begin{equation}
m_{kk'}=\frac{\sum_{i \in \Upsilon(k)} \sum_{j  \in \Upsilon(k')}
m_{ij}a_{ij}}{E_{kk'}},
\end{equation}
where $E_{kk'}$ stands for the number of edges between those degree
classes (two times that number if $k=k'$). The multiplicity matrix $m_{kk'}$ is defined in the range $[0,m^c_{kk'}]$, where
$m^c_{kk'}=min(k,k')-1$ and it represents a measure of dyadic
clustering that gives a more detailed description than $\bar{c}(k)$ on how triangles
are shared among vertices of different degrees. Furthermore, as we
shall see in the following paper, it contains relevant information to
analyze the percolation properties of clustered networks.

Now, it is possible to find a relation between multiplicity and
clustering. Taking into account the fact that the
joint degree distribution can be defined as $P(k,k')= \lim_{N
\rightarrow \infty} E_{kk'}/2E$, with $E$ the total number of edges
in the network, we obtain the following closure condition at the
class level
\begin{equation}
\sum_{k'} m_{kk'}P(k,k') = k(k-1) \bar{c}(k) \frac{P(k)}{\langle k
\rangle}. \label{db}
\end{equation}
Let us emphasize that this equation is an identity fulfilled by any
network, which ties it with the degree detailed balance condition
derived in~\cite{Boguna:2002}. These identities are important
because, given their universal nature, they can be used to derive
properties of networks regardless their specific details. As an
example, in Ref.~\cite{Boguna:2003} the detailed balance condition
was used to prove the absence of epidemic threshold in scale-free
networks.

\begin{table}[t]
\begin{center} \caption{Empirical values of the average multiplicity $\bar{m}$,
the maximum multiplicity $m_{max}$, and the clustering coefficient
$C$ for different real networks.}\vspace{0.3cm}
\begin{tabular*}{0.4\textwidth}{@{\extracolsep{\fill}}|l|r|r|r|}
\hline $\hspace{0.2cm}$Network$\hspace{0.2cm}$&$\bar{m}\hspace{0.3cm}$&$m_{max}\hspace{0.2cm}$&$C\hspace{0.3cm}$\\
\hline
\hline $\hspace{0.2cm}$PIN$\hspace{0.2cm}$&$0.30\hspace{0.2cm}$ &$10\hspace{0.3cm}$ &$0.12\hspace{0.3cm}$ \\
\hline $\hspace{0.2cm}$AS$\hspace{0.2cm}$&$2.55\hspace{0.2cm}$ &$537\hspace{0.3cm}$ &$0.45\hspace{0.3cm}$ \\
\hline $\hspace{0.2cm}$PGP$\hspace{0.2cm}$&$3.31\hspace{0.2cm}$ &$94\hspace{0.3cm}$ &$0.50\hspace{0.3cm}$ \\
\hline $\hspace{0.2cm}$E-mail$\hspace{0.2cm}$&$3.90\hspace{0.2cm}$ &$28\hspace{0.3cm}$ &$0.27\hspace{0.3cm}$ \\
\hline $\hspace{0.2cm}$Coauthors$\hspace{0.2cm}$&$4.22\hspace{0.2cm}$ &$74\hspace{0.3cm}$ &$0.74\hspace{0.3cm}$ \\
\hline $\hspace{0.2cm}$WTW$\hspace{0.2cm}$&$27.13\hspace{0.2cm}$ &$163\hspace{0.3cm}$ &$0.66\hspace{0.3cm}$ \\
\hline
\end{tabular*}
\end{center}
\label{tm}
\end{table}

A global scalar measure can also be defined for dyadic clustering.
It is the average multiplicity of the network, obtained by averaging
$m_{kk'}$ over all degree classes,
\begin{equation}
\bar{m}=\sum_{k}\sum_{k'} m_{kk'}P(k,k')=\frac{\langle k (k-1)
\bar{c}(k) \rangle}{\langle k \rangle}.
\end{equation}
Values of $\bar{m}$ close to zero mean that there are no triangles.
When $\bar{m} \lesssim 1$, triangles are mostly disjoint and their
number can be approximated as $T(k)\lesssim k/2$. Otherwise, when
$\bar{m} \gg 1$, triangles jam into edges, with many triangles
sharing the same edge. Table~I shows empirical values for the
average multiplicity $\bar{m}$, the maximum multiplicity $m_{max}$
and the clustering coefficient $C$ for different real networks.
These are the Internet at the autonomous system level
(AS)~\cite{RomusVespasbook}, the protein interaction network of the
yeast {\it S. Cerevisiae} (PIN)~\cite{Jeong:2001}, an
intra-university e-mail network~\cite{Guimera:2003}, the web of
trust of PGP~\cite{Boguna:2004b}, the network of co-authorship among
academics~\cite{Newman:2001a,Newman:2001aa}, and the world trade web
(WTW) of trade relationships among countries~\cite{Serrano:2003}. In
all cases, except for the PIN network, the value of $\bar{m}$
indicates a noticeable jamming of triangles into edges.

An alternative way to quantify dyadic clustering is by means of the
edge clustering coefficient, defined in~\cite{Radicchi:2004} as
\begin{equation}
\bar{c}(k,k')=\frac{m_{kk'}}{m^c_{kk'}}.
\label{clusteringedge}
\end{equation}
The advantage of using the normalized version $\bar{c}(k,k')$
instead of $m_{kk'}$ is that the edge clustering coefficient admits
a probabilistic interpretation. Indeed, the one-vertex clustering
coefficient $\bar{c}(k)$ can be viewed as the probability that two
neighbors of a vertex of degree $k$ are connected. $\bar{c}(k,k')$
can in its place be interpreted as the probability that an edge
connecting two vertices of degrees $k$ and $k'$ share a common
neighbor.

\section{Triadic clustering}
\label{Sec:Q}
\begin{figure}[t]
\epsfig{file=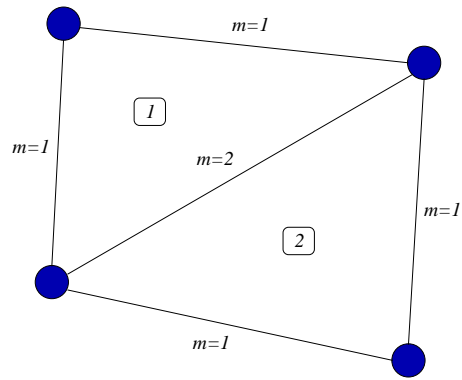,width=6cm}
 \caption{This figure illustrates the different information encoded
 in the functions $P(k,k')$ and $Q(k,k')$. In this simple graph,
 there are two kind of vertices, two of them with degree $k=2$ and
 the other two of degree $k=3$. Each edge is labeled with its
 multiplicity. The function $(2-\delta_{k,k'})P(k,k')$ tells us
 which is the probability that a randomly chosen edge connects two
 vertices of degrees $k$ and $k'$, respectively. Then, the
 probability that a randomly chosen edge connects two vertices of
 degrees $k=2$ and $k=3$ is $4/5$ and the probability of connecting
 two vertices of degree $k=3$ is $1/5$. In the case of $Q(k,k')$, we
 first have to chose randomly a triangle --either triangle labeled 1
 or 2 in this example-- and from this triangle one of its edges is
 randomly chosen. Using this procedure, the probability that an edge
 connects two vertices of degrees $k=2$ and $k=3$ is $2/3$ that
 corresponds to first chose one triangle --with probability $1/2$
 each in this particular graph-- and then one edge --with
 probability $1/3$. Analogously, the probability of connecting two
 vertices of degree $k=3$ is $1/3$. Then we can write that
 $P(2,3)=P(3,2)=2/5$, $P(3,3)=1/5$ and $Q(2,3)=Q(3,2)=2/6$,
 $Q(3,3)=2/6$.} \label{fig3}
\end{figure}

Clustering is a measure of three point correlations, although it is
not evident from the definitions of one-vertex clustering and dyadic
clustering, respectively calculated as $\bar{c}(k)$ and $m_{kk'}$.
To clarify this point, we use a similar approach to the one followed
when analyzing two-point correlations. In that case, we made use of the
matrix $E_{kk'}$, which counts the number of edges among different
degree classes, to define the joint degree distribution $P(k,k')$
giving information on the probability that a randomly chosen edge of
the network is connecting two vertices of degrees $k$ and $k'$. In
the case of triadic clustering, the fundamental object is not any
more the edge but the triangle itself. Thus, let us define a
completely symmetric tensor $T_{kk'k''}$, which measures the number
of triangles connecting vertices of the degree classes $k$, $k'$ and
$k''$ when $k \ne k' \ne k''$, two times the number of triangles
when two of the indices are equal, and six times the number of
triangles when the three indices are equal. This tensor satisfies
the following identity
\begin{equation}
\sum_{k'}\sum_{k''} T_{kk'k''}=\sum_{i\in \Upsilon (k)} 2 T_{i}=k(k-1)\bar{c}(k)P(k)N.
\end{equation}
Then, we can define a joint distribution
\begin{equation}
Q(k,k',k'') \equiv \frac{T_{kk'k''}}{\langle k \rangle \bar{m} N}
\end{equation}
which measures the probability that a randomly chosen triangle
connects three vertices of degrees $k$, $k'$, and $k''$. The one
point marginal distribution is in this case
\begin{equation}
Q(k)=\sum_{k'}\sum_{k''} Q(k,k',k') =\frac{k(k-1)\bar{c}(k)P(k)}{\langle k \rangle \bar{m}}.
\label{eq:Q(k)}
\end{equation}
The two-point marginal distribution $Q(k,k')=\sum_{k''} Q(k,k',k'')$ has an interesting interpretation. Indeed, it measures properties of the degrees of connected
vertices and, in this sense, it is similar to $P(k,k')$.  The main
difference between both distributions is the way in which edges are
selected. In the case of $P(k,k')$, an edge is randomly chosen and
then one asks for the degrees at the ends of such edge. This
selection mechanism implies that all edges in the network have the
same probability to be chosen. In the case of $Q(k,k')$, one first
selects a triangle with uniform probability among all the triangles
present in the network and, once the triangle has been selected, one
of its edges is randomly chosen. Then, the degrees of the vertices
attached to this edge are measured. If one edge is shared by more
than one triangle, this edge will be selected more often than edges
that do not participate in triangles or in just one triangle.  This
implies that, in this case, edges are chosen with a non uniform
probability which is proportional to their multiplicity
(Fig.~\ref{fig3} sketches this selection mechanism). This allows us
to write
\begin{equation}
Q(k,k')=\frac{m_{kk'}P(k,k')}{\bar{m}}.
\label{eq:Q(k,k')}
\end{equation}
Equations (\ref{eq:Q(k)}) and (\ref{eq:Q(k,k')}) eventually complete
the fundamental functions which describes transitivity properties in complex
networks. Indeed, being clustering a property that involves three
distinct vertices, the most complete description is given by the
function $Q(k,k',k'')$ which refers to triadic clustering.
Nevertheless, when no that much information is required, we can work
with the two-vertices marginal distribution $Q(k,k')$. However, by
doing so, a new quantity encoding the kernel of dyadic clustering,
the multiplicity $m_{kk'}$, naturally appears accounting for the
fact that edges can participate in more than one triangle. If we are
interested in single vertices only, we are lead to the one-vertex
marginal distribution $Q(k)$ which, again, introduces in a natural
way the concept of clustering coefficient $c(k)$. All this means
that, in fact, the functions $c(k)$ and $m_{kk'}$ are just
projections of the same fundamental object, described by
$Q(k,k',k'')$.

Dealing in practice with the three-variables function $Q(k,k',k'')$
when studying triadic clustering is a rather complex task. A
practical way to quantify triadic clustering requires the
introduction of a new measure. To this end we propose to quantify
the average multiplicity of edges among nearest neighbors in
triangles attached to a vertex of degree $k$, which we call average
nearest neighbors multiplicity $\bar{m}_{nn}(k)$ by analogy with the
average nearest neighbors degree, the function
$\bar{k}_{nn}(k)$~\cite{Pastor-Satorras:2001}. To compute $\bar{m}_{nn}(k)$ in a
formal way, we first define the transition probability
\begin{equation}
Q(k',k''|k)=\frac{\langle k \rangle \bar{m} Q(k,k',k'')}{k(k-1)\bar{c}(k)P(k)},
\end{equation}
from where we can write
\begin{equation}
\bar{m}_{nn}(k)=\sum_{k',k''} m_{k'k''} Q(k',k''|k).
\end{equation}
As in the case of $\bar{k}_{nn}(k)$, in absence of
correlations among the degrees of vertices forming triangles, the
function $Q(k',k''|k)$ is independent of $k$, and so
will be the case for the average nearest neighbors multiplicity.
Therefore, any non-trivial dependence of $\bar{m}_{nn}(k)$ on $k$
will signal the presence of correlations between the three degrees
of the nodes that form triangles.

\begin{figure}[t]
\epsfig{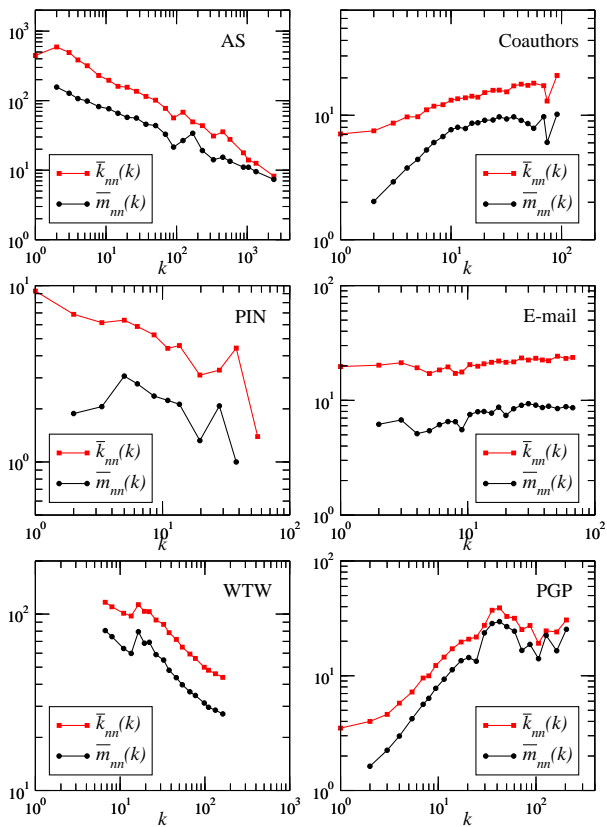}
 \caption{Empirical measures of the average nearest neighbors multiplicity as a function of $k$ compared with the average nearest
neighbors degree.} \label{m_nn}
\end{figure}
In Fig.~\ref{m_nn} we show measures of this function for the
different real networks analyzed through the paper. As one can see,
the patterns follow closely those for the average nearest neighbors
degree, that is, networks with assortative degree mixing also show
an increasing $\bar{m}_{nn}(k)$, whereas disassortative ones show
decreasing dependencies as a function of $k$. This can be intuitively understood
if we consider that the function $\bar{k}_{nn}(k)$ appears to be an
upper bound of $\bar{m}_{nn}(k)$. Despite this similarity, we also find
differences in the behavior of $\bar{m}_{nn}(k)$ as compared to
$\bar{k}_{nn}(k)$. In the case of the Internet at the autonomous
system level, we find that both functions follow a power law decay
as a function of $k$ but clearly with different exponents. In the
case of the protein interaction network, $\bar{m}_{nn}(k)$ is approximately
constant whereas $\bar{k}_{nn}(k)$ is a decreasing function of $k$.

\section{Effects of degree-degree correlations on clustering}
\label{Sec:corr}
Degree-degree correlations constrain the maximum level of
clustering a network can reach. A naive explanation for this is
that, if the neighbors of a given node have all of them a small
degree, the number of connected neighbors (and hence, the clustering
of such a node) will be bounded. This is the main idea behind the new
measure of clustering introduced in~\cite{Soffer:2005}. However, we
can make a step forward and quantify analytically this effect. The
key point is to realize that the multiplicity matrix satisfies the
inequality
\begin{equation}
m_{kk'} \le \min(k,k')-1, \label{ineq:1}
\end{equation}
which comes from the fact that the degrees of the nodes at the ends
of an edge determine the maximum number of triangles this edge can
hold. Multiplying this inequality by $P(k,k')$ and summing over $k'$
we get
\begin{equation}
k(k-1) \bar{c}(k) \frac{P(k)}{\langle k \rangle} \le \sum_{k'}
\min(k,k')P(k,k')-\frac{kP(k)}{\langle k \rangle},
\end{equation}
where we have used the closure condition Eq.~(\ref{db}). This inequality, in
turn, can be rewritten as
\begin{equation}
\bar{c}(k) \le 1-\frac{1}{k-1} \sum_{k'=1}^{k} (k-k') P(k'|k) \equiv
\lambda(k). \label{inequality}
\end{equation}
Notice that $\lambda(k)$ is always in the interval $[0,1]$ and,
therefore, $\bar{c}(k)$ is always bounded by a function smaller (or
equal) than $1$. In the limit of very large values of $k$,
Eq.~(\ref{inequality}) reads
\begin{equation}
\bar{c}(k) \le \lambda(k) \approx \frac{\bar{k}_{nn}^{r}(k)-1}{k-1}
\label{inequality_approx}
\end{equation}
where $\bar{k}_{nn}^{r}(k)$ is the average nearest neighbors degree
of a vertex with degree $k$. The superscript $r$ (of reduced) refers
to the fact that it is evaluated only up to $k$ and, therefore,
$\bar{k}_{nn}^{r}(k) \le k$. For strongly assortative networks
$\bar{k}_{nn}^{r}(k) \sim k$, so that $\lambda(k) \sim {\cal O}(1)$
and there is no restriction in the decay of $\bar{c}(k)$. In the
opposite case of disassortative networks, the sum term in the right
hand side of Eq.~(\ref{inequality}) may be fairly large and then the
clustering coefficient will have to decay accordingly.

It is important to mention that, although $\lambda(k)$ is an upper
bound of $\bar{c}(k)$, it is not the lowest upper bound. In fact, in
the inequality Eq.~(\ref{ineq:1}) we are not considering that the
neighbors of the two vertices of degrees $k$ and $k'$ might have not
enough free connections. An obvious case corresponds to vertices of
degree $1$. If the vertex of degree $k$ has $N(1|k)$ neighbors of
degree $1$ (others than the one of degree $k'$) and the one of
degree $k'$ has $N(1|k')$ neighbors of degree $1$, then, the
corrected inequality would be
\begin{equation}
m_{kk'} \le \min(k-N(1|k),k'-N(1|k'))-1. \label{ineq:2}
\end{equation}
The problem is that, now, $N(1|k)$ is an stochastic quantity with
expected value $\langle N(1|k) \rangle = (k-1)P(1|k)$ which, again,
depends on the mixing properties of the network. This contribution
is important in networks with a large number of vertices of degree
$1$.

\begin{figure}[h]
\epsfig{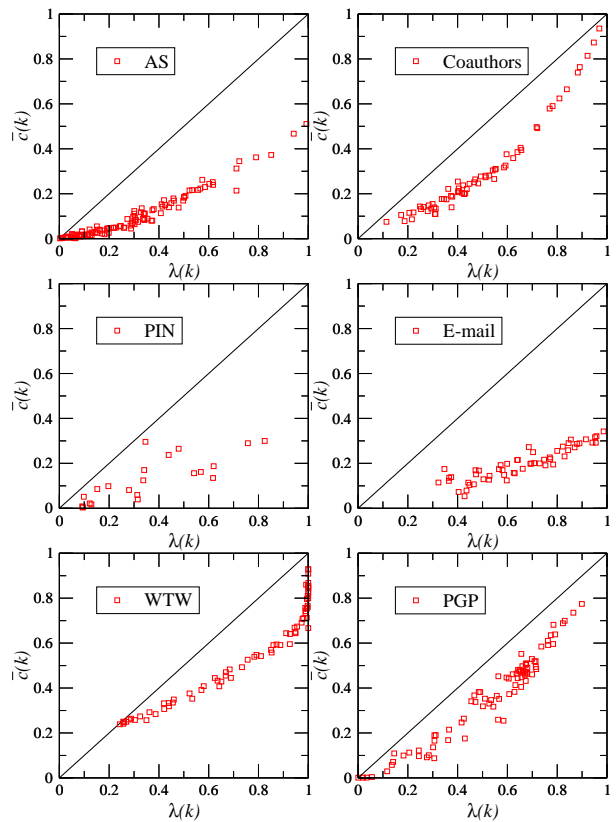}
 \caption{Clustering $\bar{c}(k)$ versus the maximum value
$\lambda(k)$ for several real networks. In all cases, empirical
measures fall below the diagonal line, validating the inequality
Eq.~(\ref{inequality}).} \label{inequality2}
\end{figure}
\begin{figure}[t]
\epsfig{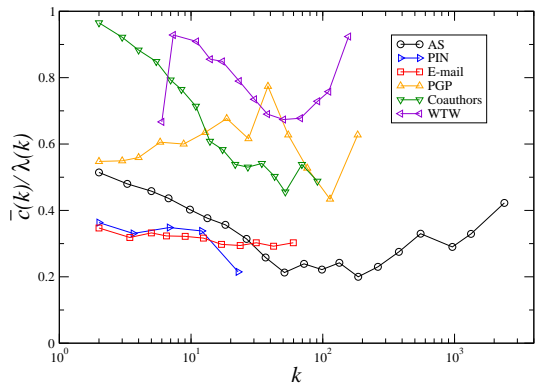}
 \caption{Empirical measures of the ratio between the clustering coefficient $\bar{c}(k)$ and the maximum value $\lambda(k)$ for different real networks.} \label{constraint_realnets}
\end{figure}

\begin{figure}[t]
\begin{tabular}{cc}
\epsfig{file=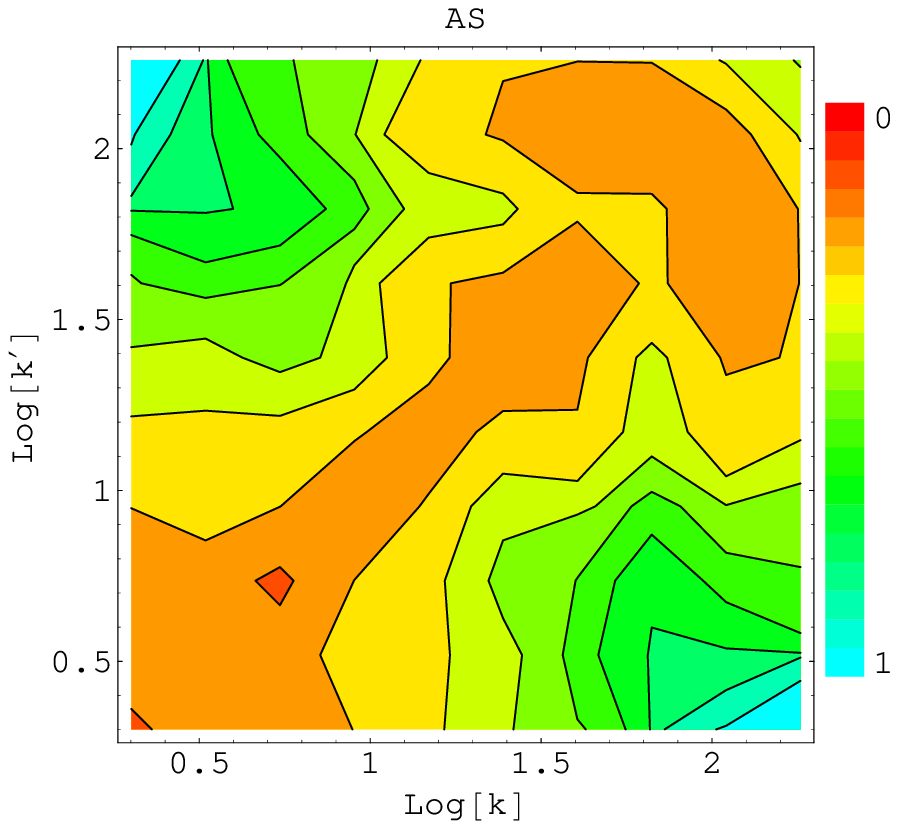,width=4.3cm} & \epsfig{file=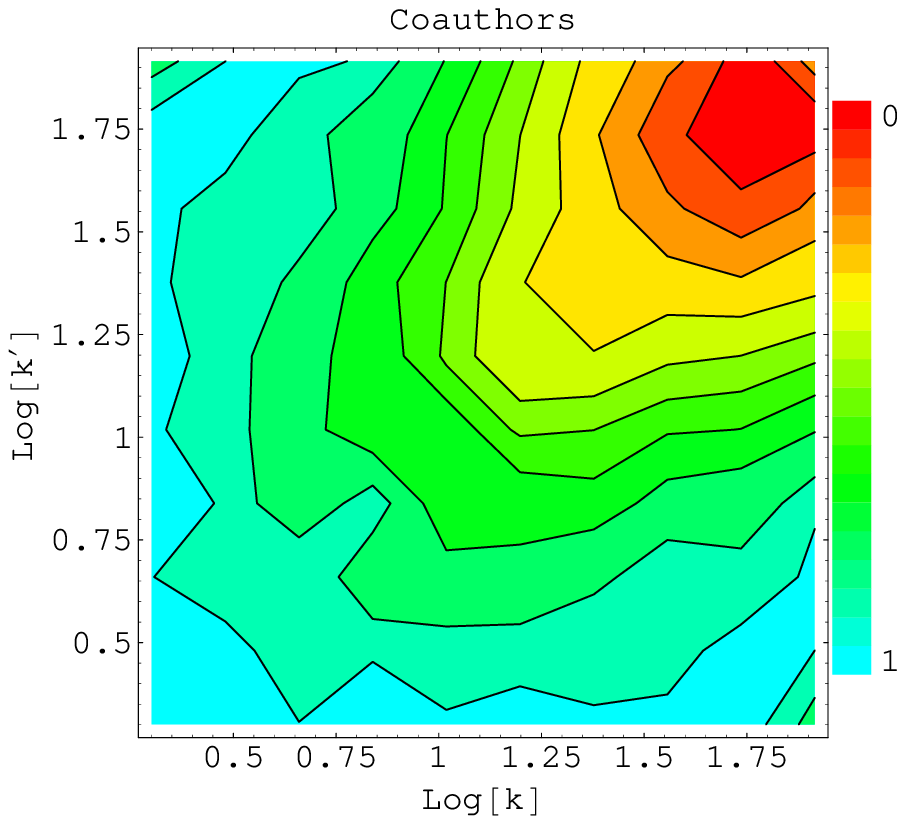,width=4.3cm}\\
\epsfig{file=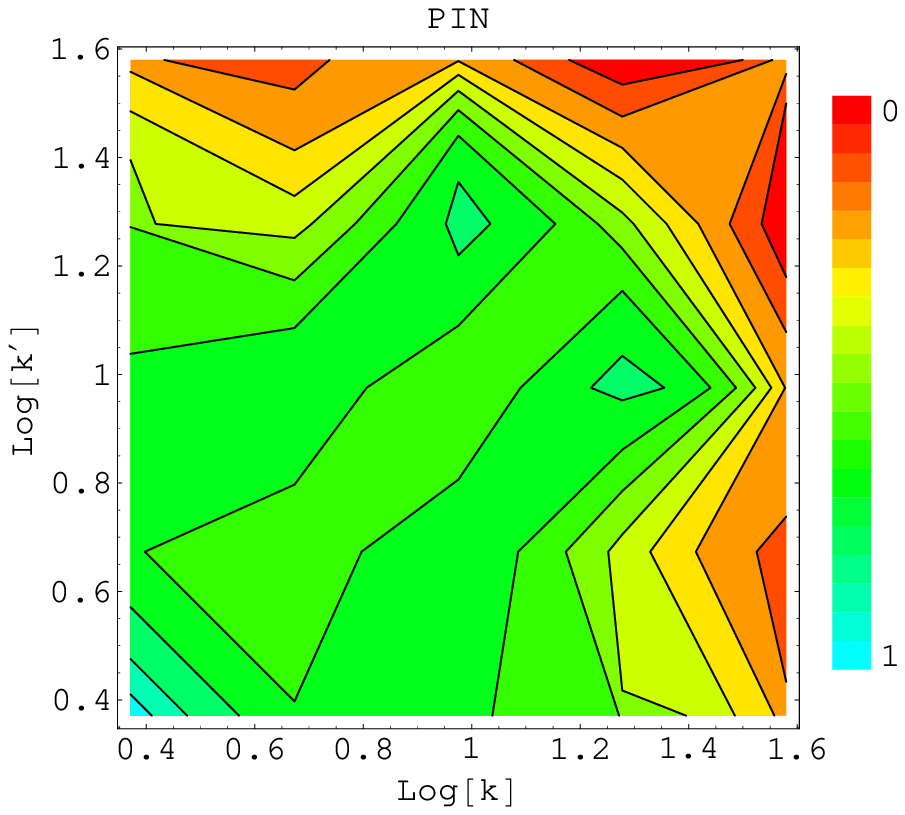,width=4.3cm} & \epsfig{file=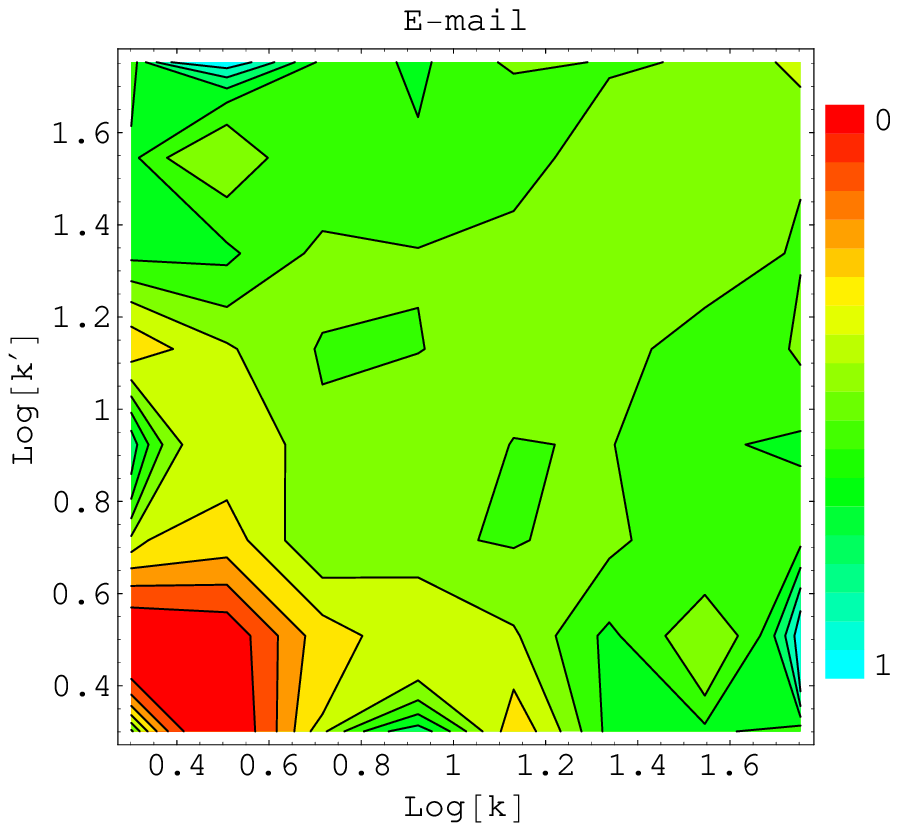,width=4.3cm} \\
\epsfig{file=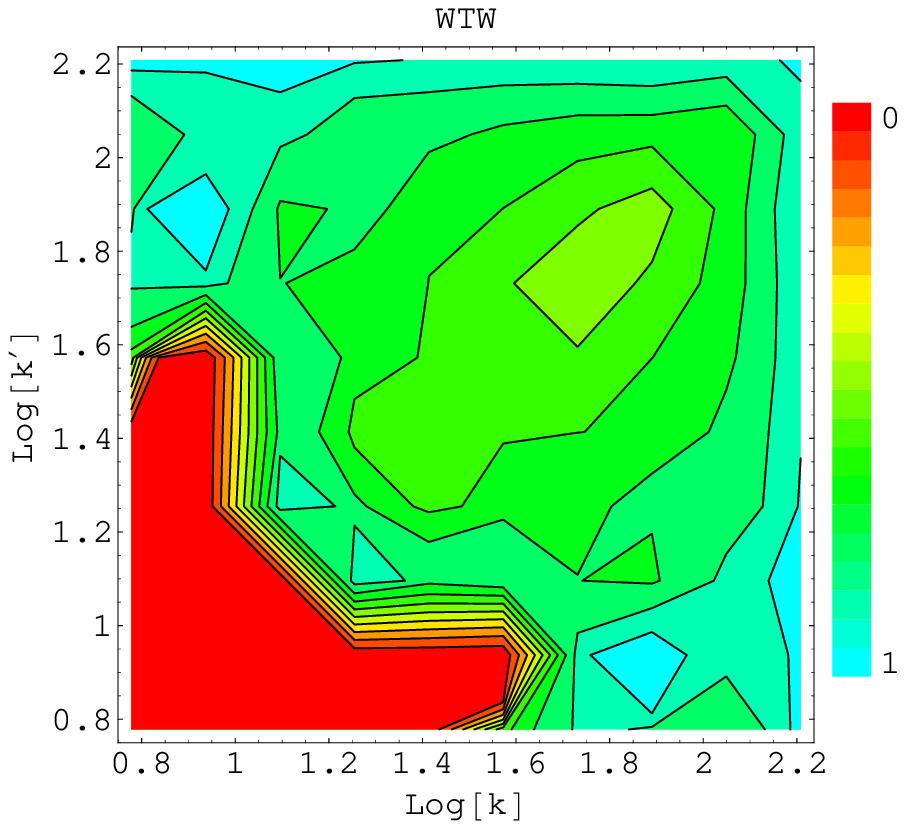,width=4.3cm} & \epsfig{file=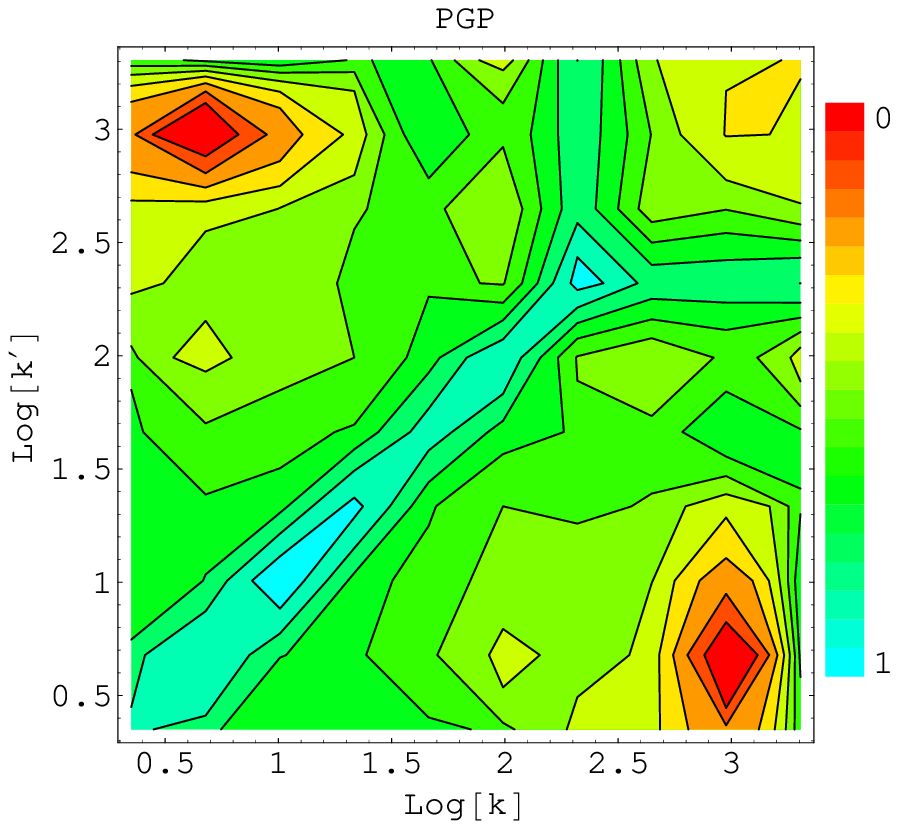,width=4.3cm}
\end{tabular}
 \caption{Contour plots of the edge clustering coefficient
$\bar{c}(k,k')$ as a function of $k$ and $k'$ for the different real
networks analyzed.} \label{edgeclustering}
\end{figure}

The interplay between degree correlations and clustering can also be
observed in real networks. We have measured the functions
$\lambda(k)$ and $\bar{c}(k)$ for several empirical data sets,
finding that the inequality Eq.~(\ref{inequality}) is always
satisfied. In Fig.~\ref{inequality2}, we plot the clustering coefficient
$\bar{c}(k)$ as a function of $\lambda(k)$. Each dot in these
figures corresponds to a different degree class. As clearly seen, in
all cases the empirical measures lie below the diagonal line, which
indicates that the inequality Eq.~(\ref{inequality}) is always
preserved. In Fig.~\ref{constraint_realnets}, we show the ratio
$\bar{c}(k)/\lambda(k)$. The rate of variation of this fraction is
small and, thus, the degree dependent clustering coefficient can be
computed as $\bar{c}(k)=\lambda(k) f(k)$, where $f(k)$ is a slowly
varying function of $k$ that, in many cases, can be fitted by a
logarithmic function. This result implies that, to a large extent,
the functional dependence of $\bar{c}(k)$ is given by the particular
shape of the degree-degree correlations. On the other hand, this
also suggests that the edge clustering coefficient, given by
Eq.~(\ref{clusteringedge}), is also a weakly dependent --if not
independent-- function of the degrees $k$ and $k'$. Indeed,
empirical measures of $\bar{c}(k,k')$ in the studied real networks
support this conjecture. Figure \ref{edgeclustering} shows contour
plots of $\bar{c}(k,k')$ using a logarithmic binning of the axes. In
all cases, there is a dominant color, which indicates that the edge
clustering is approximately constant. As expected from the information shown in
Fig.~\ref{constraint_realnets}, the AS network and the Coauthor
network are the less constant, although the variation of
$\bar{c}(k,k')$ across different $(k,k')$ domains is not very
pronounced. This result is particularly important because, unlike to what happens for the degree dependent clustering coefficient $c(k)$, it allows to
approximate dyadic clustering in many cases by a constant value.

\subsection{Scale-free networks}
Scale-free networks with degree distributions of the form $P(k)\sim k^{-\gamma}$ belong to a special class of networks which
deserve a separate discussion. Indeed, it has been shown that, when
the exponent of the degree distribution lies in the interval $\gamma
\in (2,3]$ and its domain extends beyond values that scale as
$N^{1/2}$, disassortative correlations are unavoidable for high
degrees~\cite{Park:2003,Burda:2003,Boguna:2004,Catanzaro:2005}.
Almost all real scale-free networks fulfill these conditions and,
hence, it is important to analyze how these negative correlations
constrain the behavior of the clustering coefficient. Let us assume
a power law decay of the average nearest neighbors degree of the
form $\bar{k}_{nn}(k)\sim \kappa k^{-\delta}$. One can prove that
this function diverges in the limit of very large networks as
$\bar{k}_{nn}(k)\sim \langle k^2 \rangle \sim k_{c}^{3-\gamma}$,
where $k_{c}$ is the maximum degree of the
network~\cite{Boguna:2003}. Then, the prefactor $\kappa$ must scale
in the same way which, in turn, implies that the reduced average
nearest neighbors degree behaves as
\begin{equation}
\bar{k}_{nn}^r(k)\sim  k^{3-\gamma-\delta}.
\end{equation}
Then, from Eq.~(\ref{inequality_approx}) the exponent of the degree
dependent clustering coefficient, $\alpha$, must verify the
following inequality
\begin{equation}
\alpha \ge \gamma+\delta-2.
\end{equation}
Just as an example, in the case of the Internet at the Autonomous
System level~\cite{Vazquez:2002}, the reported values for these
three exponents ($\alpha=0.75$, $\gamma=2.1$, and $\delta=0.5$)
satisfy this inequality close to the limit ($\alpha=0.75 \ge
\gamma+\delta-2=0.6$).

\section{Uncorrelated networks and the distinction between weak and strong clustering}
\label{Sec:uncorr}
When analyzing two-point correlations, the notion of uncorrelated
network corresponds to a network in which the joint distribution
$P(k,k')$ factorizes as
\begin{equation}
P(k,k')=\frac{k k' P(k) P(k')}{\langle k \rangle ^2}.
\end{equation}
In the context of triangles, a network is uncorrelated when
\begin{equation}
Q(k,k',k'')=Q(k)Q(k')Q(k''),
\end{equation}
where $Q(k)$ was given in Eq.~(\ref{eq:Q(k)}). The question is
whether functions $P(k,k')$ and $Q(k,k',k'')$ can factorize
simultaneously. First, we restrict to the study of the factorization
of $Q(k,k')$ instead of $Q(k,k',k'')$ since, due to the symmetric
attribute as a tensor of the last function, the factorization of
$Q(k,k')$ is a sufficient condition for the factorization of
$Q(k,k',k'')$. Indeed, the function $Q(k,k')$ measures correlations
between connected vertices when edges are weighted by their
multiplicity, whereas $P(k,k')$ measures these correlations when
edges are chosen with uniform probability. Given this difference in the selection mechanism of edges, $Q(k,k')$ and
$P(k,k')$ cannot factorize simultaneously when the sample of edges
is highly heterogeneous in their multiplicity values. In contrast, when
$m_{ij}$ is either $0$ or $1$, the sample of edges corresponding to
triangles will become homogeneous and whenever $Q(k,k')$ factorizes,
$P(k,k')$ factorizes too (for degrees larger than 1). In this case,
we can write
\begin{equation}
m_{kk'}\propto(k-1)(k'-1)\bar{c}(k) \bar{c}(k').
\end{equation}
Since in this approach $m_{kk'} \le 1$ for $\forall k,k'$, we have
that
\begin{equation}
 \bar{c}(k) \le \frac{1}{k-1}
\end{equation}
for any uncorrelated network at the two-vertex level. In other situations, one can construct
uncorrelated networks at the level of triangles but, at the same
time, there will appear some correlations at the level of edges and vice-versa.

This suggests to partition the space of clustered networks into
two main categories: {\it weak transitivity} --for networks with $
\bar{c}(k) \le (k-1)^{-1}, \forall k$-- and {\it strong transitivity} in the
opposite case. As we will show in the following paper~\cite{following}, the percolation
properties of clustered networks are totally different depending on
which one of these categories the network belongs to. This is
related to the fact that, in the strong transitivity regime, the
overlap of triangles is important, favoring thus the emergence of
subgraphs which are tightly interconnected, the so-called $k$-cores~\cite{Dorogovtsev:2006}. In contrast, in the {\it weak transitivity} class, triangles are mostly disjoint and the topological properties of such networks are close to that of  unclustered ones.

\section{Conclusions}
\label{Sec:conclusions}
In this paper, we have provided a new and powerful formalism to understand transitive relations in complex networks. We have defined a new fundamental quantity, $Q(k,k',k'')$, which measures the probability that a randomly chosen triangle connects three vertices of degrees $k$, $k'$, and $k''$. The summation over one variable of this fundamental distribution gives information about two of the vertices participating in the triangle and, in a natural way, introduces the multiplicity of edges among two classes of degrees $k$ and $k'$, $m_{kk'}$. The summation of $Q(k,k',k'')$ over two of its variables gives information about the properties of vertices that participate in triangles and, as in the previous case, naturally defines the degree-dependent clustering coefficient $\bar{c}(k)$. To quantify the extent of the correlations encoded in $Q(k,k',k'')$, we have proposed a new metric, the average nearest neighbors multiplicity $\bar{m}_{nn}(k)$, finding interesting patterns when measured in real networks. We have also found that, in real networks, the edge clustering coefficient, defined as the ratio between $m_{kk'}$ and $min(k-1,k'-1)$, is a weakly dependent function of the degrees $k$ and $k'$. This could serve as a basis for modeling of clustered networks. This result also suggest that the functional form of the degree-dependent clustering coefficient is mainly determined by the two-vertex correlation structure of the network. Last but not least, we have found the conditions for the simultaneous factorization of $Q(k,k',k'')$ and $P(k,k')$. This is only possible if $\bar{c}(k)<(k-1)^{-1}$. This partitions the space of clustered networks into two main categories, networks with {\it weak transitivity} --those that satisfy $\bar{c}(k)<(k-1)^{-1}$-- and networks with {\it strong transitivity} --those that do not. In the first class, the multiplicity of edges is either zero or one and triangles are disjoint. In the second class, edges
are forced to share many triangles, giving rise to highly interconnected subgraphs. We shall see in the following paper how the class a network belongs to changes its percolation properties.

\begin{acknowledgments}
This work has been
partially supported by DGES, Grant No. FIS2004-05923-CO2-02 and
Generalitat de Catalunya Grant No. SGR00889. M. B. thanks the School
of Informatics at Indiana University, where part of this work was
developed.
\end{acknowledgments}


\end{document}